\magnification=\magstep1
\line{\hskip 4.4in UCR HEP-T184\hfill}
\centerline{\bf Is It Possible to Disentangle an Entangled Quantum State?}
\vskip .3cm
\centerline{Shu Yuan Chu}
\centerline{Physics Department, University of California, Riverside, California 92521}
\vskip .4cm
Experimental tests of the suggestion that the generalization of Wheeleer and Feynman's time 
symmetric system is the dynamical basis underlying quantum mechanics are considered.  In a
time-symmetric system, the instantaneous correlations exhibited by two spatially separated
particles in an entangled state can be established through other particles, and can reveal
advanced interaction effects.  In particular, the existence of advanced gravity waves may be
detectable through suitable arrangements at the Laser Interferometer Gravitation-Wave 
Observatory.
\vfill\eject

One of the most fundamental predictions of quantum mechanics is the correlations between
particles in an entangled quantum state in an Einstein-Podolsky-Rosen type experiment.  The
concept of causality, that cause must precede its effect, is one of the most fundamental
principles in physics, firmly established by innumerable observations.  It is therefore a
truly surprising result, although perhaps not generally perceived as such, that Wheeler
and Feynman were able to show that the time-symmetric electrodynamics, where the advanced 
interaction is equal in strength to the causal retarded interaction, is a viable alternative
to the usual theory containing only the retarded interaction [1].  By imposing the complete
absorber boundary condition, which requires that there are enough absorbers to absorb all
the radiation in the system, they demonstrated that the effect of the advanced 
electromagnetic interaction on a test charge is completely canceled, except for the 
radiation reaction on an accelerated one.  In their time-symmetric theory, the advanced
interaction is the crucial ingredient for producing the radiation reaction, which is the
only manifestation of the advanced interaction identified by Wheeler and Feynman.  Because
the radiation reaction is required in the usual theory by energy considerations, it 
cannot be taken as decisive evidence for the existence of advanced interaction.

With the benefit of hindsight we must say that it is unfortunate that Wheeler and 
Feynman did not pursue other possible manifestations of the advanced interaction.  The
reason is what one might loosely denote as the ``quantum non-locality" problem, which
was finally formulated precisely by Bell in 1964 [2].  Bell showed that quantum mechanics
demands the violation of Bell's inequality, which implies that there are instantaneous 
correlations between particles separated by nonzero spatial distances.  These 
instantaneous correlations on the surface violate the combined requirement of causality
and the relativity principle that interactions propagate with a finite maximum velocity.
This difficulty can be overcome in the time-symmetric system.

When both retarded and advanced interactions are present, instantaneous correlations 
between two spatially separated particles can be established through other particles.
The simplest way is to have a third paticle, which correlates with one of the two
particles through the advanced interaction and with the other one through the retarded
interaction.  As long as it takes an equal amount of time for the interaction to
travel between the third particle and each of the two particles in question, 
instantaneous correlations will be established.  More complicated ways involving a
larger number of particles are also possible provided that the correlations established
are instantaneous.  This last condition singles out a small fraction of the total number
of particles available in the system.

If Wheeler and Feynman had pursued the possible manifestations of the advanced 
interaction further, it would have been likely for them to conclude that there are
instantaneous correlations between two particles in their time-symmetric system.
Because only a small fraction of the particles in the system are involved in 
establishing these correlations, bypassing the complete absorber condition, the
effect of the advanced interaction is not completely canceled in this situation.
If the possible existence of instantaneous correlations had been suggested as a 
prediction of the time-symmetric electrodynamcis in the late 1940s, long before
Bell's insightful analysis, the subsequent experimental detection [3] of such
correlations would then be considered as the verification of the prediction of the
time-symmetric system rather than that of quantum mechanics.  The fact that,
quantum mechanics, later on, also was found to predict such instantaneous 
correlations through the violations of Bell's inequality, would only provide a
strong argument for the suggestion that the time symmetric system is a possible
dynamic basis underlying quantum mechanics [4]. 

Since it is impossible to change the historical sequence of events, how can we test
whether the time-symmetric system is a possible dynamic basis of quantum mechanics
at this late date?  One way is to search for manifestations of the advanced 
interaction not predicted by quantum mechanics already.  An obvious area to
consider is gravity, where the complete quantum mechanical theory is not yet available.

When Wheeler and Feynman proposed their time-symmetric electrodynamics, Einstein remarked
that he saw nothing wrong with their proposal but did not know how to apply it to
gravity [5].  In their proposal, there are only charged particles interacting 
directly among each other with a finite velocity of propagation for the interactions.
It is this distinct feature of direct interaction that demands the existence of the
advanced interaction with strength equal to that of the retarded interaction.  The
electromagnetic field emerges as an ``adjunct field" that summarizes all the 
information about the motion of the charged particles.  The recent development of
string theories, capable of unifying gravity with other interactions, raises the
expectation that the natural generalization of the theory of Wheeler and Feynman
from point particles to strings may lead to a time-symmetric theory of gravity [6].
The gravitational field will then also emerge as an ``adjunct" field summarizing the
behavior of the strings.

As an initial step to choose between the time-symmetric approach and the conventional
causal theory, we would like to examine whether there is any experimental evidence
against the existence of advanced gravitational interaction.  If it is not precluded
by existing experimental results, then we would like to explore the possibility of
detecting the advanced gravitational interaction experimentally.  In the final
analysis this is how we can decide which approach is the better choice without being
influenced by any preconceived preferences.

In any experiments involving only a static gravitational field, one cannot distinguish
between the advanced or the retarded choices.  All five classic tests of general
relativity: the gravitational redshift of spectral lines, the deflection of light
by the sun, the precession of the perihelia of the orbits of the inner planets, the
time delay of radar echoes passing the sun, and the precession of a gyroscope in orbit
around the earth, to a good approximation are carried out in a static field [7].  Staying
within the experimental capabilities available in the near future, we conclude that 
only gravity-wave experiments are left to be examined.

The only existing indirect evidence of gravity waves is the famous discovery of
Taylor and Hulse [8].  Their observation of the ever shortening separation between
a whirling pair of neutron stars measures the radiation reaction of the gravity
waves radiated by the binary system.  As mentioned above, in Wheeler and Feynman's
electrodynamics, the radiation reaction depends critically on the existence of
the advanced interaction.  Hence, the experimental results of Taylor and Hulse
do not provide any evidence against the existence of advanced gravity waves.  Next,
let us consider experiments that directly detect the gravity waves.

In order to detect the advanced gravity waves, the effect of the advanced gravitational
interaction on the detector clearly must not be completely canceled.  Wheeler and 
Feynman analyzed the possibility of incomplete cancellation of the effect of the
advanced interaction in general, and they concluded that there is no logical 
inconsistency [1].  The complete cancellation of the effect of the advanced 
interaction on a test charge in the electromagnetic case, depends on the complete
absorber condition {\it and} the fact that the retarded field is a solution of Maxwell's
homogeneous equations [9].  This latter condition is not satisfied in the gravity
case.  It is therefore possible that the effect of the advanced gravitational 
interaction is not completely canceled, and there are detectable advanced gravity
waves.

How can one detect these advanced gravity waves?  The most unambiguous signature of
advanced gravity waves in a time-symmetric theory is the detection of correlated
signals separated by time intervals equal to 2T, where T is the time of travel
between the source and the detector of the gravity wave.   Since the distances
between the detectable sources of gravity waves and the earth are expected to be
billions of light years, this is clearly not a practical way to proceed.

If the location of the gravity-wave source is known, two or more detectors placed at 
different distances from the source can distinguish between the advanced and the
retarded waves.  The retarded wave, which is outgoing from the source, will reach
the detector nearest to the source before it reaches the detectors farther away.
The advanced wave, which is incoming, will reverse the order of the sequence of events.

Perhaps the most exciting opportunity to detect the advanced gravity waves from 
sources of unknown location will be at the Laser Interferometer Gravitational Wave
Observatory, or LIGO, where multiple detectors are expected to be operational at
different locations around the globe in the first decade of the next century [10].
From the details of the signals seen by the detectors, consisting of the outputs 
of their photodiodes, it will be possible to deduce the location of the source
and the waveforms of the gravity wave for each of the two polarizations [11].  These 
capabilities should make it possible to decide whether the signal corresponds to a
retarded wave or an advanced wave.  For instance, consider the gravity-wave signal 
from the coalescence of two black holes.  The gravity waveforms of the initial 
inspiral phase and final ringdown phase of the coalescing process are understood 
from the solutions of the Einstein field equation (those from the middle 
coalescence phase are not yet understood).  The waveforms from the inspiral phase
are expected to oscillate with gradually growing amplitude and frequency, while
those from the ringdown phase to oscillate with fixed frequency and gradually
dying amplitude [11].  If the signal corresponding to the inspiral phase arrives
before that from the ringdown phase, it is a retarded wave.  If the order of
arrival of the signals is reversed, it is an advanced wave.  The point to be
stressed here is that the signals from the advanced waves should be comparable
in strength to those from the conventional retarded waves except for the reversed
order of arrivals, therefore equally detectable by LIGO.

The experimental detection of the advanced gravity waves is a crucial test of the
suggestion that the time-symmeteric system is the dynamical basis underlying quantum
mechanics.  We look forward to the time when the debate about the solution to the
``quantum measurement" problem can be resolved experimentally rather than philosophically.

Acknowledgment: It is a great pleasure to thank Professors F.Cummings, B.R.Desai, S.Y.Fung,
J.G.Layter, J.G.Nickel,B.C.Shen, G.J.Van Dalen, and C.H.Woo for many helpful discussions.
\vskip 1cm
\noindent References

\noindent  1. J.A.Wheeler and R.P.Feynman, Rev.Mod.Phys. {\bf 21},425 (1949). 

\noindent  2. J.S.Bell, Physics, {\bf 1}, 195 (1964).

\noindent  3. A.Aspect, J.Dalibard and G. Roger, Phys. Rev. Lett. {\bf 49},1804 (1982).

\noindent  4. J.G.Cramer, Rev.Mod.Phys.{\bf 58},647 (1986); Int.J.Theor.Phys.{\bf 27},
227(1988). S.Y.Chu, Phys. Rev. Lett. {\bf 71}, 2847 (1993).

\noindent  5. The remark was quoted in R.P.Feynman, {\it Surely You're Joking, Mr. Feynman!}
(Norton, 1985),p.80.

\noindent  6. The development of a time-symmetric theory of gravity was not pursued in the
early effort at generalization: M.Kalb and P.Ramond, Phys.Rev.{\bf D9},2273.  For some 
preliminary results of this development see S.Y.Chu, gr-qc/9802070.

\noindent  7. S. Weinberg,{\it Gravitation and Cosmology} (Wiley, 1972), Chapter 8.

\noindent  8. R.A.Hulse and J.H.Taylor, Astrophys.J.{\bf 195},L51 (1975).  J.H.Taylor and
J.M.Weisberg, Astrophys.J.{\bf 345}, 434 (1989).

\noindent  9. R.P.Feynman, Phys.Rev. {\bf 74},939 (1948),footnote No. 6.

\noindent 10. A.Abramovici et al., Science {\bf 256}, 325 (1992).

\noindent 11. For detailed descriptions, see K.S.Thorne, {\it Black Holes and Time
Warps} (Norton, 1994),pp.378-396.

\end